\documentclass[a4paper,11pt]{article}
\usepackage{float}
\usepackage{jheppub} 
\usepackage{verbatim}
\usepackage[compat=1.1.0]{tikz-feynman}

\newcommand{\OL}[1]{ \hspace{2pt}\overline{\hspace{-2pt}#1
   \hspace{-1pt}}\hspace{1pt} }
\newcommand{\kten}{{\kappa_{10}}}
\newcommand{\calR}{{\cal R}}

\newcommand{\tilF}{{\tilde F}}
\newcommand{\barG}{{\OL G}}

\newcommand{\leaveout}[1]{}

\usepackage[normalem]{ulem}
\usepackage{slashed}
\usepackage{bigints}
\usepackage{braket}
\usepackage{cleveref}
\usepackage{comment}
\usepackage{bm}		    
\usepackage{xcolor}
\usepackage{csquotes}

\newcommand{\bea}{\begin{eqnarray}}
\newcommand{\eea}{\end{eqnarray}}
\newcommand{\bg}{\begin{eqnarray}}
\newcommand{\eg}{\end{eqnarray}}

\title{Quadratic Axion Couplings in String Theory}
\preprint{USTC-ICTS/PCFT-26-47}

\author[a]{Naman Agarwal}
\author[a,b]{Andrew R.~Frey,}
\author[c,d]{Ratul Mahanta,}
\author[b]{and Evan McDonough}

\affiliation[a]{Department of Physics \& Astronomy, \\
University of Manitoba, Winnipeg, Manitoba R3T 2N2, Canada}
\affiliation[b]{Department of Physics,  and Winnipeg Institute for Theoretical Physics,\\ 
University of Winnipeg, Winnipeg MB, R3B 2E9, Canada}
\affiliation[c]{Interdisciplinary Center for Theoretical Study, University of Science and Technology of China, Hefei, Anhui 230026, China}
\affiliation[d]{Peng Huanwu Center for Fundamental Theory, Hefei, Anhui 230026, China}

  \emailAdd{agarwaln@myumanitoba.ca}
  \emailAdd{a.frey@uwinnipeg.ca}
  \emailAdd{ratul.mahanta@ustc.edu.cn}
  \emailAdd{e.mcdonough@uwinnipeg.ca}

\abstract{
Axions and axion-like particles are a compelling candidate for physics beyond the standard model. While many axion searches are focused on the linear coupling to photons $\theta F \tilde{F}$, the possibility of a quadratic coupling to the electromagnetic kinetic term, $\theta^2 F^2$, leads to novel phenomenology and new opportunities for testing axion-like particles. In this work we propose mechanisms for generating this coupling in string theory, which can be broadly classified as classical, perturbative, and non-perturbative. In benchmark examples, we find that both perturbative and non-perturbative quantum contributions such as instantons lead to couplings that are suppressed, $g \ll 1$ in units of $1/f^2$ where $f$ is axion decay constant, though easily larger than analogous coupling of the QCD axion that is generated through loops of charged pions. These analyses suggest that quadratic axion couplings to gauge fields are ubiquitous in string theory, and should be taken seriously as a probe of the string theory axiverse, both of string theory candidates for dynamical axions, such as dark matter or dark energy, and for spectroscopy of the string theory axiverse.}

\begin{document}
\maketitle
\flushbottom

\section{Introduction}
\label{sec:intro}
A signature prediction of string theory is the existence of a spectrum of axion-like particles at low energies \cite{Arvanitaki:2009fg,Cicoli:2012sz}.\footnote{A spectrum of axion-like particles is also predicted by certain field theory models \cite{Alexander:2024nvi,Alexander:2023wgk,Maleknejad:2022gyf}.} The axions could range in mass over many orders of magnitude \cite{Broeckel:2021dpz, Ling:2025nlw}, and may have been responsible for inflation, dark matter, dark energy, and more. A significant experimental program is dedicated to searching for these axions \cite{ParticleDataGroup:2022pth}.

The primary focus of work to date, theoretical and experimental searches alike, is the conventional coupling of axions to photons: $\theta F \tilde{F}$, where $\theta$ is the axion, $F$ is the Maxwell field strength tensor, and $\tilde{F}$ is the Hodge dual of $F$. An alternative approach is to search for axions through a hypothetical coupling $\theta^2 F^2$ that is quadratic in the axion. This coupling leads to new and novel phenomenology. For example, if $\theta$ is coherent scalar field that is evolving in time, such as axion dark matter, this implies a variation of the fine-structure constant that can be tested with atomic clocks \cite{Arvanitaki:2014faa,Beadle:2023flm}. Perturbations to the field imply localized variations of fundamental constants, which can be probed by pulsar timing arrays \cite{Gan:2025icr} and by gravitational wave detectors such as LISA \cite{Gan:2026vph}. Even if $\theta$ has no background value, and does not constitute the observed dark matter, the $\theta^2 F^2$ coupling can enable $\theta$ particles to be be pair produced in regions of strong inhomogeneous magnetic fields, such as around pulsars \cite{Prabhu}, allowing for {\it axiverse spectroscopy}, with all light axion-like particles being produced by the pulsar. 
The quadratic axion coupling may also play a role during cosmic inflation, e.g.~in generation of primordial magnetic fields \cite{Martin:2007ue}.  More generally, quadratic axion couplings to matter can modify the local dark matter density profile and axion mass, with implications for fifth-force experiments and searches for violations of the equivalence principle \cite{Bouley:2026frx,Brzeminski:2026pgz}.

The quadratic axion coupling to photons is the focus of this work. Concretely, we define the interaction of interest to be
\begin{equation}
\label{eq:gaagg}
    {\cal L}_{\theta\theta \gamma \gamma} = g_{\theta\theta \gamma\gamma} \hat\theta^2 \hat F^2 ,
\end{equation}
where $\hat\theta$ is the (dimensionful) canonically normalized axion and $\hat F$ is the field strength tensor of the canonically normalized electromagnetic field, i.e., defined by ${\cal L}_{\rm EM}=-\frac{1}{4}\hat F^2$. The coupling $g_{\theta\theta\gamma \gamma}$ has dimension of inverse mass squared, and can be expected to scale as roughly $g_{\theta\theta\gamma \gamma} \sim 1/f^2$ where $f$ is the axion decay constant. This intuition holds true for the QCD axion: Ref. \cite{Beadle:2023flm} finds  the result
\begin{equation}
\label{eq:gaagg1loop}
g_{\theta\theta\gamma\gamma}^{\rm 1-loop}\simeq \frac{\alpha}{16\pi^2f^2}\simeq 4.6 \times 10^{-5} \frac{1}{f^2} 
\end{equation}
generated by loops of charged pions. 

In this work we undertake the first survey of quadratic axion couplings in string theory. We consider classical, perturbative, and non-perturbative contributions, and consider broad possibilities for the mass and identity of the candidate axion and of the gauge field to which it couples. Among many possibilities, we identify three paths to generate quadratic axion couplings within the context of the well studied Type IIB axiverse: classical couplings that appear in the 10-dimensional and dimensionally reduced supergravity actions, perturbative corrections from integrating out heavy moduli fields (which can also be framed as a classical backreaction effect), and non-perturbative corrections due to ED3 instantons or gaugino condensation on D7 branes. While different in detail, versions of these mechanisms will also apply in other string compactifications (such as intersecting brane models in type IIA string theory).

Our results can be summarized as follows:
\begin{itemize}
    \item[] {\bf I. Classical couplings}: The 10D supergravity action contains couplings of the schematic form $(C F)^2$, where $C$ is a $p$-form potential and $F$ a field strength. In a general compactification, these terms reduce to a $\theta^2 F^2$ term in the lower-dimensional effective action. The DBI action on a brane includes similar terms. Additionally, the real part of the chiral coordinates of the 4D supergravity (for a CY orientifold compactification) combine the K\"ahler moduli and 2-form axions, which can introduce the couplings of interest.
    
    \item[] {\bf II. Perturbative Corrections:} While string loop corrections do not generate axion couplings to $F^2$, there are nonetheless perturbative corrections from integrating out heavy moduli in the low energy effective field theory. Applied to the $\theta_b$ axion of large volume scenario (LVS) compactifications, we find an exponential suppression at large volume, $g_{\theta \theta \gamma \gamma} \propto e^{ - a {\cal V}^{2/3}}$. This suppression can be mitigated by a modest ${\cal V}$ and a small $a$, leading to couplings that are significantly larger than the benchmark from charged pion loops. Integrating out a heavy modulus at tree level is equivalent to determining the classical backreaction of the axion on the ground state of the heavy modulus. In addition, a triangle anomaly in the 4D supergravity may introduce quadratic axion couplings in some cases.
    \item[] {\bf III. Non-Perturbative Corrections:} The gauge kinetic function receives non-perturbative corrections in much the same way as the superpotential. We find that both gaugino condensation and instantons generate a coupling that is suppressed at large volume like $g \propto e^{ - a {\cal V}^{2/3}}$, similar to the coupling from integrating out heavy moduli. Again depending on parameter values, this can be large or small.
\end{itemize}
In summary, we find many mechanisms in the axiverse to realize the quadratic axion coupling, Eq.~\eqref{eq:gaagg}.

These results demonstrate the quadratic axion coupling to gauge field kinetic energy is ubiquitous in string theory. However this comes with several caveats: We do not perform detailed particle physics model building, and do not demand that gauge field in question be the $U(1)_{\rm EM}$ of the Standard Model of particle physics, or even that the gauge group be $U(1)$. We do not specialize to the case of axion dark matter or even ultralight axions. Instead, we perform a scan over mechanisms by which axions in string theory can be coupled to gauge fields in string theory through the desired interaction. With an eye towards the broad phenomenology, from the early universe to direct detection, we keep an open mind towards the axion and gauge field of interest, and perform a broad, if not exhaustive, scan over possibilities. We embed a set of general mechanisms into string theory models and estimate the magnitude of the coupling. The aim is to guide future studies which will take a more targeted approach at testing the axiverse through the quadratic axion coupling.

The structure of this paper is as follows: In Sec.~\ref{sec:origins} we perform a scan over mechanisms to generate the quadratic axion coupling, Eq.~\eqref{eq:gaagg}, classifying them as classical, perturbative, or non-perturbative. In Sec.~\ref{sec:models}, we embed a number of these into concrete string theory models, in both the construction of \cite{Kachru:2003aw} (KKLT) and LVS frameworks for moduli stabilization, and estimate the size of the coupling in each case. We conclude in Sec.~\ref{sec:discussion} with a discussion of directions for future work.


\section{Quadratic Axion Couplings In String Theory}
\label{sec:origins}


Quadratic axion couplings to gauge fields are ubiquitous in string theory. They emerge from the classical structure of string theory, from quantum corrections to the four-dimensional effective action, and from non-perturbative effects such as instantons. These can all be captured by the gauge kinetic function of the gauge field, $f_{\rm gauge}$, which determines the kinetic terms for gauge fields as 
\begin{align}
    \mathcal{L}_{\rm gauge, kin} = -\frac{M_p^2}{4}\mathrm{Re}(f_{\rm gauge})F_{\mu\nu}F^{\mu\nu}-\frac{M_p^2}{4}\mathrm{Im}(f_{\rm gauge})F_{\mu\nu}\tilde{F}^{\mu \nu}
    \label{eq:gaugekin}
\end{align}
 where $\tilde{F}_{\mu\nu}$ denotes the Hodge dual of the field strength tensor, and we assume an Abelian gauge field for simplicity. The imaginary part of the gauge kinetic coupling encodes the canonical axion-like coupling to gauge fields, $\theta F \tilde{F}$. The real part of the the gauge kinetic function encodes the couplings to $F^2$, such as the axion $\theta^2 F^2$ coupling that is the focus of this work.   In general the gauge kinetic function can be expressed as (see, e.g., Refs.~\cite{Blumenhagen:2006xt, Akerblom:2007nh, Blumenhagen:2008ji, Blumenhagen:2009qh}),
\begin{equation}
\label{eq:fterms}
    f= f_{\rm classical} + f_{\rm pert} + f_{\rm non-pert}
\end{equation}
corresponding to the classical, perturbative, and non-perturbative contributions.  Here we consider a general taxonomy of contributions, and in Sec.~\ref{sec:models} we consider explicit examples in the context of compactification on a Calabi-Yau (CY) orientifold.

For the sake of generality we keep an open mind to the many species of axion-like particles that emerge from string theory, such as the fundamental axion $C_0$ and axions that arise through dimensional reduction.  We work in the context of Type IIB string theory, where the 4-dimensional axion fields are given by
\begin{equation}
    b^a = \int_{\Sigma_a} B_2 \ , \qquad   c^a =  \int_{\Sigma_a} C_2 \ ,  \qquad \theta^\alpha = \int_{D_\alpha} C_4\ ,  \label{IIBaxions}
\end{equation}
where $B_2$ is the Kalb-Ramond $2$-form, $C_2$ and $C_4$ are the Ramond-Ramond $2$- and $4$-form fields, and $\Sigma_a$ and $D_\alpha$ denote a basis of $2$-cycles and $4$-cycles respectively, enumerated by the Hodge numbers of the internal manifold $\mathcal{X}$ with $a=1,..,h^{1,1}_-(\mathcal{X})$ and $\alpha=1,..,h^{1,1}_+(\mathcal{X})$. 

We also keep an open mind as to the gauge field of interest. Gauge fields are ubiquitous in four-dimensional effective field theories arising from string theory, arising from branes and from dimensional reduction. This leads to spectrum of visible and hidden $U(1)$ gauge bosons, the string theory {\it photoverse} \cite{Coudarchet:2025dfd}. In this work we do not pursue particle physics model building or enforce {\it a priori} that the model realizes the $U(1)$ electromagnetism of the Standard Model of particle physics. We consider consider both Abelian and non-Abelian gauge fields.

\subsection{Classical Contributions}

We first consider classical contributions to the gauge kinetic function. These are readily apparent even without specifying a compactification, e.g., from the action for Type IIB supergravity, which is given by 
\begin{eqnarray}
S_{\rm IIB} &=& {1\over 2\kten^2} \int d^{10} x \sqrt{-g_{\rm s}}\Biggl\{
e^{-2\phi}
\left[\calR_{\rm s} + 4(\nabla \phi)^2 \right] - {F_{1}^2\over 2} -
{1\over 2 \cdot 3!} G_{3} \wedge \star \barG_{3} -
{\tilF_{5}^2\over 4\cdot5!} \Biggr\} \nonumber\\
&& + {i\over 8  \kten^2}
\int e^{\phi}{C_{4}\wedge
G_{3}\wedge\barG_{3}}\, \ ,
\label{IIBS}
\end{eqnarray}
where $G_{3}= F_{3} - i S H_{3}$, $S=1/g_s - i C_{0}$ is the axiodilaton\footnote{Following the conventions of \cite{Cicoli:2023opf}.}, and ${\tilde F}_{5} $ is a corrected field strength given by
\begin{equation}
{\tilde F}_{5} = F_{5} - {1\over 2} C_{2}\wedge H_{3} + {1\over 2}
B_{2}\wedge F_{3}\ .
\end{equation}
Dimensional reduction to four dimensions of the the term $G_3 \cdot \barG_3$ generates a $\theta^2 F^2$-type coupling of the $C_0$ axion to the $U(1)$ vector field descending from dimensional reduction of $B_2$. Similarly, the $\tilde{F} _5 ^2$ term generates an $\theta^2 F^2$-type coupling of the $c^a$ axions to the gauge fields descending from $B_2$, and of the $b^a$ axions coupled to the gauge fields descending from $C_2$. These are all absent in a Calabi-Yau manifold, where there are no 1-cycles, but are otherwise present in the four-dimensional theory. 
Similar considerations apply in compactifications of the type IIA string.

Tree-level couplings also emerge from D-branes, coming from the Dirac-Born-Infeld action, which on a Dp-brane is given by
\begin{equation}
S_{\rm Dp, DBI} = -  \int {\rm }d^{p+1} \sigma \, T_p \sqrt{ -  {\rm det} (h_{\mu \nu} + \partial_\mu \Phi^a \partial_\nu \Phi_a  + F_{\mu\nu} + B_{\mu \nu }}) .
\end{equation}
Expansion of the determinant generates the familiar Yang-Mills theory with kinetic term $F_{\mu \nu} F^{\mu \nu}$, along with interactions such as $B_{\lambda\rho}B^{\lambda\rho}F_{\mu \nu} F^{\mu \nu}$. Dimensional reduction leads to an effective  coupling $b^2 F^2$ for the $B_2$ axion coupled to the gauge field on the brane worldvolume. Similarly, dimensional reduction of an NS5 brane, the action for which follows from S-duality of the D5-brane, generates an quadratic coupling for the $C_2$ axion. These couplings have been computed and studied in the context of axion monodromy inflation in e.g., \cite{Barnaby:2011qe}.

Tree-level couplings also emerge in the context of widely studied Calabi-Yau orientifold compactifications. Indeed, the chiral coordinates of the ${\cal N}=1$ supergravity effective theory are given by \cite{Grimm:2004uq} as 
\begin{align}
    &G^a =\bar{S} b^a + i c^a = \frac{b^a}{g_s}+i(c^a-C_0 b^a) \,,\qquad \tau_\alpha = \frac12\,k_{\alpha\beta\gamma}\,t^\beta t^\gamma\,,\\
    &T_\alpha = \tau_\alpha + i \theta_\alpha - \frac14\, g_s k_{\alpha a b} G^a (G+\bar{G})^b\,.
\label{ChiralCoord}
\end{align}
where $\tau_\alpha$ denote 4-cycle volumes, $t^\alpha$ denote 2-cycle volumes, and $k_{\alpha \beta \gamma}$ denotes the triple-intersection numbers of the manifold. Consider the simple case of a manifold with $h^{1,1}=2$ and $h^{1,1}_+=h^{1,1}_-=1$,  with a divisor $D_1$ whose orientifold image is $D_2$. One can then define the orientifold-even 4-cycle $D_+ \equiv  D_1 \cup D_2$ and and orientifold-odd 4-cycle $D_-\equiv D_1 \cup (-D_2)$. The K\"ahler modulus then takes the simple form
\begin{equation}
T = \tau + i \theta -\frac14 g_s k G(G+\bar{G})\,,
\label{chiralCoordinates}
\end{equation}
where $k_{+--}\equiv k$.

Now consider a D7 brane in this geometry. The gauge kinetic function, including two-form axions but assuming vanishing worldvolume fluxes for simplicity, is given by \cite{Grimm:2011dj} as
\begin{equation}
\label{eq:fD7}
    f_{\rm D7} = T ,
\end{equation}
where $T$ is the chiral coordinate on the 4-cycle wrapped by the brane. Consider for example a D7 brane wrapping the cycle $D_+$ in the simple example outline above, following Refs.~\cite{Grimm:2011dj,Cicoli:2023qri}. The real part of the gauge kinetic function is simply 

\begin{equation}
    f_{\rm D7} = \tau - \frac{k}{2g_s}b^2 .
\end{equation}
From equation Eq.~\eqref{eq:gaugekin}, this implies a coupling $b^2 F^2$ with coupling constant $g_{bb\gamma\gamma}\propto-k/(g_s)$ if the internal volume $\tau$ is stabilized rather than the chiral coordinate $T$. Note that $k<0$ to avoid the $c_2$ axion from being a ghost \cite{Cicoli:2023qri}.
The extension to the case with worldvolume fluxes is given in \cite{Cicoli:2023qri}.

\subsection{Perturbative Contributions}

We now turn to perturbative quantum contributions to the quadratic axion couplings to gauge fields, corresponding to the second term in Eq.~\eqref{eq:fterms}.  

\begin{figure}[t!]
\includegraphics[width=\textwidth]{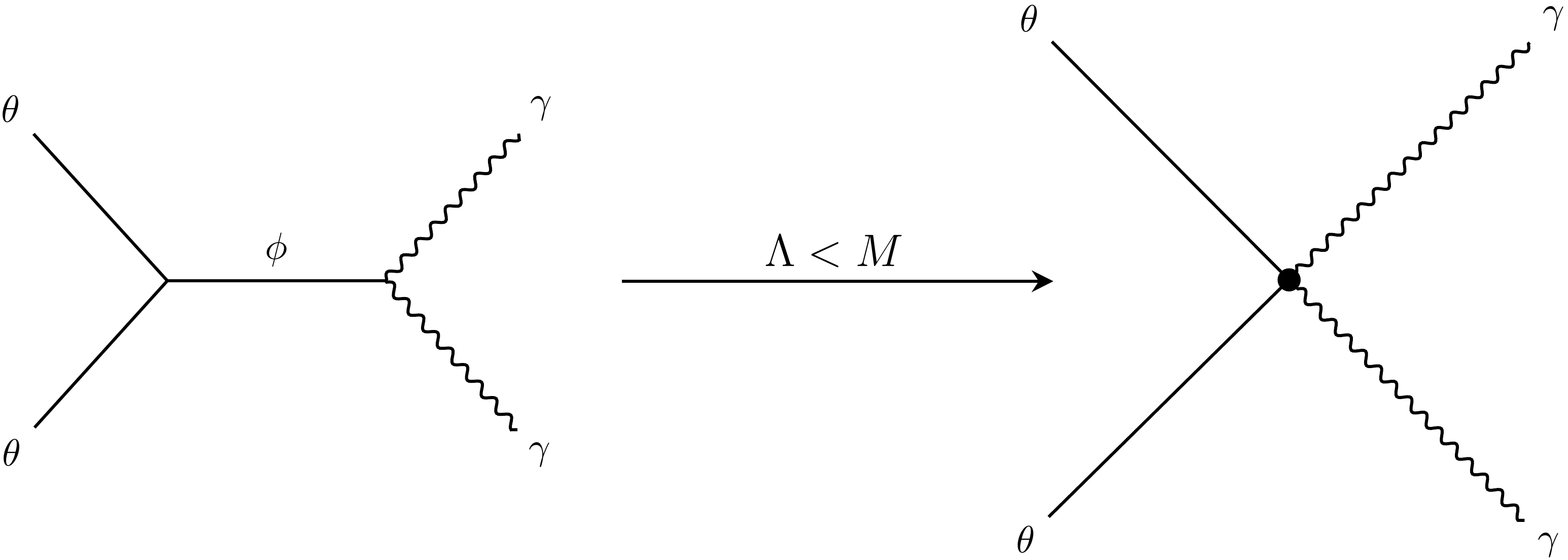}
\caption{Effective axion quadratic coupling obtained by integrating out the modulus in the EFT if the mass of the modulus $M$ is higher than the cutoff scale $\Lambda$.}
 \label{fig: Feynman}
\end{figure}

Quantum corrections in the four dimensional effective field theory, arising from integrating out heavy volume moduli, easily generate such couplings. For example, consider a toy model 
\begin{equation}
    {\cal L} =  - \frac{1}{2}(\partial \phi)^2 - \frac{1}{2}(\partial \theta)^2 - \frac{1}{g^2(\phi)} F_{\mu \nu} F^{\mu \nu} - \frac{1}{2} M^2 \phi^2 -  m(\phi)^2 f^2 \left[ 1 - \cos\left( \frac{\theta}{f}\right)\right] ,
\end{equation}
where $\phi$ and $\theta$ respectively represent a heavy modulus field of mass $M$ and a light axion field with $\phi$-dependent mass set by $m(\phi)$, which depends on the heavy field. For simplicity take $m^2(\phi)= m_0^2(1 + \lambda \phi)$ and $1/g^2 = 1/g_0^2 +a\phi$, which implies interactions $ m_0^2 \lambda \phi \theta^2$ and $a \phi F^2$. Integrating out $\phi$ for $M$ larger than the cutoff $\Lambda\gg m_0$ leads to an effective interaction $\sim (a\lambda m_0^2/M^2) \theta^2 F_{\mu \nu}F^{\mu \nu}$ between the axion and the photon, e.g. from tree-level $s$-channel $\phi$ exchange (in fact, the interaction is of the form $\cos(\theta/f) F^2$ for this model) as illustrated in  Fig.~\ref{fig: Feynman}. This toy model structure is realized in string theory (with additional higher-dimension terms) via the classical gauge coupling ${\rm Re}(f_{\rm D7})=\tau$ and the non-perturbative superpotential $W= A e^{- a T}$ leading to the axion potential $V_{\rm axion}\sim e^{-a \tau} \cos(\theta)$, which couples the two fields.

Integrating out heavy fields at tree level can be described equivalently as a classical backreaction effect. 
In the path integral formalism, we can carry out tree-level Gaussian integral of $\phi$ by completing squares, but an equivalent approach is to shift the integration variable by the classical value for $\phi$, which is $\approx -(aF^2/4+m_0^2\lambda\theta^2/2)/M^2$ for a heavy modulus (and the same shift of the field when completing the square). Then the gauge kinetic energy, axion potential, and modulus mass terms yield a total interaction $(a\lambda m_0^2/8M^2)\theta^2 F^2$. 
Either style of argument also applies to models with multiple moduli including a nontrivial moduli space metric. 
Because we can frame the calculation in terms of classical backreaction, it also describes the coupling between an oscillating background axion field and the gauge field (when the modulus is heavy).
As in the Wilsonian description above, the modulus needs to be much heavier than the axion for the classical backreaction description to hold; if the modulus is not heavier, axion motion will cause the modulus to oscillate rather than track the instantaneous $\theta$-dependent minimum.

Backreaction leads to couplings $g_{\theta\theta\gamma\gamma} \sim a\lambda f^2 (m_0/M)^2/f^2$ in terms of the axion decay constant.
This will be small compared to the decay constant unless the scales $1/a,1/\lambda \ll f$ since $m_0\ll M$ if it is consisent to integrate out the modulus.

There are also generally contributions to gauge couplings at loop level.
String loop corrections in Type IIB string theory do not directly generate axion dependence of the gauge kinetic function: the corrections depend only on the complex structure moduli \cite{Blumenhagen:2009qh} while the axions appear in the K\"ahler moduli (see also \cite{Kaplunovsky:1995jw} for heterotic, \cite{Berg:2004ek} and references therein for Type IIB, \cite{Lust:2003ky} for Type IIA and \cite{Camara:2008zk} for Type I). Note though that string loops do depend on the K\"ahler moduli in the IIA string and may therefore generate the couplings of interest.

However, loop effects can indirectly generate $\theta^2 F^2$ couplings. As explained in \cite{Kaplunovsky:1994fg,Kaplunovsky:1995jw}, anomaly cancellation necessitates Wess-Zumino terms in the supergravity action, which correct the gauge coupling at 1-loop in the form of a non-holomorphic contribution to the effective gauge coupling. The correction depends on the K\"{a}hler potential and is given by \cite{Kaplunovsky:1994fg,Kaplunovsky:1995jw,Conlon:2009xf,Conlon:2009kt,Kachru:2019dvo}
\begin{equation}
\label{eq:loop}
    \frac{1}{g^2} \supset - \frac{N}{2\pi^2 M_p^2} K(T,\bar{T})
\end{equation}
where $g$ is the physical gauge coupling, $N$ is in reference to an $SU(N)$ gauge group, and $K$ is the K\"{a}hler potential. From the chiral coordinates, Eq.~\eqref{ChiralCoord}, this generates a $b^2 F^2$ coupling (when the chiral coordinate $T$ is the stabilized modulus as opposed to $\tau$).

\subsection{Non-Perturbative Contributions}

Finally, we come to non-perturbative contributions to the gauge kinetic function. The gauge kinetic function receives non-perturbative corrections in much the same way as the superpotential. Contributions include ED3 instantons and gaugino condensation on D7 branes, which generate couplings for $C_4$ axions, and more exotic contributions such as ED(-1)-instantons or gaugino condensation on D3-branes at singularities \cite{Cicoli:2012fh}, which can generate couplings for $C_0$ axion \cite{Blumenhagen:2009qh}. Fluxed instantons can generate couplings for $c_2$ and $b_2$ axions.

There is another class of non-perturbative contributions, namely those which appear in the K\"{a}hler potential and enter the gauge coupling via the loop correction Eq.~\eqref{eq:loop}.
The K\"{a}hler potential receives non-perturbative corrections which can depend on the axions,  e.g. from ED1 instantons, \cite{Cicoli:2023qri}
\begin{equation}
K_{\rm ED1} = -3 \ln \left(\Re(T) + \frac{k}{2 g_s}\,b^2 +\sum_{\substack{n\in\mathbb{N} \\ \hat{\mathfrak{f}}_-\in\mathbb{Z}}} A_{n,\hat{\mathfrak{f}}_-} e^{-2\pi n \left(\frac{t}{\sqrt{g_s}} + k \hat{\mathfrak{f}}_- G \right)}\right),
\label{ED1}
\end{equation}
where 
\begin{equation}
t = \sqrt{\frac{2}{\tilde{k}}\left(\Re(T) + \frac{k}{2 g_s}\,b^2\right)}.
\end{equation}
This couples the $C_2$ axions to  gauge fields on D7 branes carrying worldvolume flux. Similar results also apply for gaugino condensation on D5 branes \cite{Cicoli:2023qri}.

\section{Concrete Models}
\label{sec:models}

We now embed the general mechanisms discussed in Sec.~\ref{sec:origins} into concrete models, utilizing the KKLT \cite{Kachru:2003aw} and 
LVS \cite{Balasubramanian:2005zx,Conlon:2005ki} framework for moduli stabilization. 

The action for scalars in $\mathcal{N}=1$ 4D EFT is given in terms of K\"ahler potential $K(\phi^I,\phi^{\bar{I}})$ and superpotential $W(\phi^I,\phi^{\bar{I}})$ that connect to kinetic and potential terms as\footnote{see for example \cite{Freedman:2012zz} for a review.}
\begin{align}
    \mathcal{L}_{\rm mod, kin}=-K_{I\bar{J}}\partial^\mu \phi^{I}\partial_\mu\bar{\phi}^{\bar{J}}
\end{align}
where $K_{I\bar{J}}=\partial_I \partial_{\bar{J}}K$ is the K\"ahler metric and 
\begin{align}
    V=e^{K/M_p^2}\left(K^{I\bar{J}}D_IW D_{\bar{J}}\overline{W} - \frac{3|W|^2}{M_p^2}\right).
\end{align}
The covariant derivative here is given as $ D_IW=\partial_IW + \frac{1}{M_p^2}W\partial_I K$. The kinetic term for gauge fields can be expressed as Eq.~\eqref{eq:gaugekin}. Here $M_{p}=2.435 \times 10^{18}$ GeV is the reduced Planck mass. While we use dimensionless fields (scalar and vector), we refrain from working in Planck units and keep explicit factors of $M_{p}$ to ease the later conversion of quantities to units of GeV. Canonically normalized fields are denoted with a hat, i.e., $\hat\theta$ for a canonically normalized axion.
 
\subsection{Perturbative Correction} \label{sec: PerQuantCorr}

\subsubsection{KKLT}\label{sec: KKLTbackreaction}

 We begin with the simplest possible construction: KKLT \cite{Kachru:2003aw}. Consider the 4D effective field theory for KKLT with volume modulus $\tau$ and $C_4$ axion $\theta$. The EFT is derived by dimensional reduction of Type IIB string theory compactified over a Calabi-Yau orientifold with $h^+_{1,1}=1$. We  include a nilpotent field $X$ for SUSY breaking and to get a Minkowski or dS minimum \cite{Kallosh:2014via,Aparicio:2015psl}. The K\"ahler potential is given as 
\begin{align}
    K=-3M_p^2\ln{(T+\overline{T})}+3M_p^2\frac{X \overline{X}}{T+\overline{T}}
\end{align}
where $T=\tau+i\theta$. This gives us the kinetic term for $\tau$ and $\theta$ as
\begin{align}\label{eq: KKLTKin}
    \mathcal{L}_{\rm mod, kin}=-\frac{3M_p^2}{4\tau^2}(\partial\tau)^2-\frac{3M_p^2}{4\tau^2}(\partial\theta)^2.
\end{align}
The superpotential for KKLT is 
\begin{align}
    W=W_0 + Ae^{-aT}+MX ,
\end{align}
where $W_0$ is the GVW flux superpotential \cite{Gukov:1999ya}, the exponential is a non-perturbative term (which stabilizes the volume modulus) due to gaugino condensate on D7 branes wrapping a four cycle, and $M$ is a constant which can be chosen for a Minkowksi vacuum. The scalar potential then becomes  
\begin{align}\label{eq: KKLTPotential}
   V_{\mathrm{KKLT}}=\frac{a^2A^2e^{-2a\tau}}{6\tau M_p^2}+\frac{aA^2e^{-2a\tau}}{2\tau^2 M_p^2}+\frac{aAW_0e^{-a\tau}}{2\tau^2 M_p^2}\cos{(a\theta)}+\frac{M^2}{12M_p^2\tau^2} ,
\end{align}
where $W_0<0$, $a$, and $A$ are parameters that descend from 10D theory. The Minkowski minimum of this potential is at \cite{Aparicio:2015psl}
\begin{align}
    \theta_\star&=0 , \qquad
    \tau_\star=-\frac{5}{2a}-\frac{1}{a}W_{-1}\left(\frac{3W_0}{2Ae^{5/2}}\right) \textnormal{ for}\nonumber\\
    M^2&=-\frac{9aW_0^2\left[1+2W_{-1}\left(\frac{3W_0}{2Ae^{5/2}}\right)\right]}{4W^2_{-1}\left(\frac{3W_0}{2Ae^{5/2}}\right)},
\end{align}
where $W_{-1}(x)$ denotes the secondary branch of the Lambert W function and $\star$ denotes the field value in vacuum.\footnote{For an exponentially small negative argument, the $W_{-1}$ branch is parametrically large and negative, which is the relevant regime for an exponentially small negative flux stabilized GVW superpotential $W_0$.}
 
The tree level gauge kinetic term in $\mathcal{N}=1$ SUGRA is expressed as
\begin{align}\label{eq: TreeGaugeKin}
    \mathcal{L}_{\mathrm{gau, kin}}\supset -\frac{M_p^2}{4}\tau_{\star} F_{\mu\nu}F^{\mu\nu}.
\end{align}

In order to unambiguously define the couplings we need to use canonically normalized fields. From the kinetic terms for the gauge field (equation (\ref{eq: TreeGaugeKin})), axion and the volume modulus (equation (\ref{eq: KKLTKin})), canonical normalization is defined as 
\begin{align}
    \hat{F}_{\mu\nu}=\sqrt{\tau_{\star}}M_p F_{\mu\nu}\, ; \, \hat{\theta}=\sqrt{\frac{3}{2}}\frac{M_p}{\tau_\star} \theta \, ; \,  \delta\hat{\tau}=\sqrt{\frac{3}{2}}\frac{M_p}{\tau_\star}\delta \tau. 
\end{align}
Using the axion stabilizing term from the equation (\ref{eq: KKLTPotential}), the decay constant of the axion can be expressed as
\begin{align}
    f=\frac{M_p}{a\tau_\star}\sqrt{\frac{3}{2}}.
\end{align}

From the KKLT potential in equation (\ref{eq: KKLTPotential}), we can expand the axion stabilizing term to get a $\tau \theta^2$ interaction term  
\begin{align}
    \mathcal{L}_{\tau \theta \theta}=-\frac{a^3A |W_0|(2+a\tau_\star)e^{-a\tau_\star}}{3\sqrt{6}M_p^5}\delta\hat{\tau}\hat{\theta}^2.
\end{align}
Also, from the tree level gauge kinetic term equation (\ref{eq: TreeGaugeKin}), we have the interaction 
\begin{align}
    \mathcal{L}_{\tau \gamma\gamma}=-\frac{1}{2\sqrt{6}M_p}\delta\hat{\tau}\hat{F}_{\mu\nu}\hat{F}^{\mu\nu}.
\end{align}
The coupling constant of the effective interaction due to integrating out $\delta\hat\tau$ is then
\begin{align}
    g_{\theta\theta\gamma\gamma}=\frac{a A |W_0| (2+a\tau_{\star}) e^{-a\tau_\star}}{24 M_p^4 \, m^2_\tau\, \tau_\star^2}\frac{1}{f^2}.
\end{align}
For values of the parameters in \cite{Kachru:2003aw}, $g_{\theta\theta\gamma\gamma} \sim 10^{-3}/f^2$.  

However, there is an important caveat here: as mentioned in section \ref{sec:origins}, 
there should be a mass hierarchy $m_{\tau}\gg m_{\theta}$ to integrate the modulus out consistently. In the case of KKLT, the ratio of masses of the volume modulus and the axion is
\begin{align}
    \frac{m^2_{\tau}}{m^2_\theta}=\frac{1+W_{-1}\left(\frac{3W_0}{2Ae^{5/2}}\right)}{W_{-1}\left(\frac{3W_0}{2Ae^{5/2}}\right)} \approx 1
\end{align}
for values of the parameters $A$ and $W_0$ in \cite{Kachru:2003aw}, and hence the required mass hierarchy is not present.

\subsubsection{Large Volume Scenario}\label{sec: LVSBackreaction}

We now consider the Large Volume Scenario \cite{Balasubramanian:2005zx,Cicoli:2008va}. This scenario can realize the same qualitative type of backreaction as KKLT but admits the mass hierarchy which is missing in the latter.

Consider a Swiss-cheese type Calabi-Yau manifold with two 4-cycles, one big and one small, the volumes of which are denoted by $\tau_b$ and $\tau_s$, respectively. The total volume of the CY can be written as 
\begin{align}
    \mathcal{V}=\tau_b^{3/2}-\tau_s^{3/2}
\end{align}
in string units. With a nilpotent field $X$ (used to break SUSY and get Minkowski minima), the K\"ahler potential is 
\begin{align}
    K=-2M_p^2 \ln{\left(\mathcal{V}+\frac{\hat{\xi}}{2}\right)} +M_p^2\frac{X\overline{X}}{\mathcal{V}^{2/3}}.
\end{align}
Here, the $\hat{\xi}\equiv \xi \,g_s^{-3/2}$ term pertains to the order $\alpha^{\prime3}$ correction where $\xi=-\zeta(3)\chi/2(2\pi)^3$ in terms of the CY Euler number $\chi$.

In LVS the small cycle is stabilized non-perturbatively, whereas the overall volume is fixed perturbatively. This leaves the $\theta_b$ axion massless. In order to generate a mass for $\theta_b$ we additionally consider a non-perturbative contribution on the large cycle, which can be either gaugino condensate or an ED3 instanton. The superpotential is given by
\begin{align}
     W=W_0 + A_s e^{-a_s T_s}  + A_b e^{-a_b T_b} +MX. 
\end{align}
which includes the GVW superpotential $W_0$ and the SUSY breaking nilpotent field. 

Then the scalar potential is approximately \cite{Cicoli:2023qri} 
\begin{align}\label{eq: LVSPotential}
    V_{\rm LVS}=&\frac{8a_s^2 A_s^2 e^{-2a_s\tau_s}\sqrt{\tau_s}}{3\mathcal{V}M_p^2}-\frac{4 a_s A_s\tau_s|W_0|e^{-a_s\tau_s}}{\mathcal{V}^2M_p^2}\cos{(a_s\theta_s)}+\frac{3|W_0|^2\hat{\xi}}{4\mathcal{V}^3 M_p^2}+\frac{M^2}{\mathcal{V}^{4/3}M_p^2}\nonumber\\
     &-\frac{4a_b A_b |W_0|e^{-a_b\tau_b}}{\mathcal{V}^{4/3}M_p^2}\cos{(a_b\theta_b)}
\end{align}
where we have used $W_0=-|W_0|$. The potential can minimized with the conditions $\mathcal{V}_\star\gg1$, $a_s \tau_{s\star} \gg 1$ and $\tau_{b\star}\gg\tau_{s\star}$ telling us that the axion $\theta_b$ stabilizing term in equation (\ref{eq: LVSPotential}) is exponentially suppressed in its contribution to the stabilization of $\mathcal{V}$ and $\tau_s$. The stabilized value of the fields are \cite{Aparicio:2015psl}\footnote{This assumes $\hat\xi$ is sufficiently large such that the LVS limit is realized, i.e. $a_s\tau_{\star}\sim\ln\mathcal V_{\star}$.}
\begin{equation}
    \theta_{s\star}=0, \qquad \theta_{b\star}=0 , \qquad 
    \tau_{s\star}=\left(\frac{\hat{\xi}}{2}\right)^{2/3}, \qquad \tau_{b\star}^{3/2}=\mathcal{V}_{\star}=\frac{3|W_0|\sqrt{\tau_{s\star}}}{{4A_s a_s }}e^{a_s \tau_{s\star}} 
\end{equation}
and $M$ fixed is to be $M^2=3\hat{\xi}\, |W_0|^2/8 a_s \tau_{s\star}\mathcal{V_\star}^{5/3}$ in the Minkowski minimum.

To canonically normalize the fields we expand the kinetic Lagrangian to leading order in the inverse volume, which gives
\begin{align}
    \mathcal{L}_{\rm mod, kin}\supset-\frac{M_p^2}{2}&\left(\frac{3}{2\mathcal{V}_\star^{4/3}}\partial_\mu\tau_b\partial^{\mu}\tau_b  +\frac{3}{2\mathcal{V}_\star^{4/3}}\partial_\mu \theta_b\partial^{\mu}\theta_b\right.\nonumber\\
    &\left.+\frac{3\sqrt{a_s}}{4\mathcal{V}_\star\sqrt{\ln\mathcal{V}_\star}}\partial_\mu \tau_s\partial^{\mu}\tau_s+\frac{3\sqrt{a_s}}{4\mathcal{V}_\star\sqrt{\ln\mathcal{V}_\star}}\partial_\mu \theta_s\partial^{\mu}\theta_s\right).
\end{align}
This establishes the canonical normalization to be 
\begin{align}
     \hat{\tau}_b&=\sqrt{\frac{3}{2}}\frac{M_p}{\mathcal{V}_\star^{2/3}}\tau_b, \qquad 
    \hat{\theta}_b=\sqrt{\frac{3}{2}}\frac{M_p}{\mathcal{V}_\star^{2/3}}\theta_b ,\nonumber\\
     \hat{\tau}_s&=\frac{M_p}{2}\left(\frac{9a_s}{\mathcal{V}_\star^2\ln{\mathcal{V}_\star}}\right)^{1/4}\tau_s, \qquad
    \hat{\theta}_s=\frac{M_p}{2}\left(\frac{9a_s}{\mathcal{V}_\star^2\ln{\mathcal{V}_\star}}\right)^{1/4}\theta_s. 
\end{align}
For D7 branes wrapping the big cycle, the gauge kinetic term for the brane gauge field is then
\begin{align}
    \mathcal{L}_{\mathrm{gau, kin}}\supset -\frac{M_p^2}{4}\tau_{b\star}F_{\mu\nu}F^{\mu\nu}+\cdots.
\end{align}
Using this, the canonical normalization for the gauge field  becomes
\begin{align}
    \hat{F}_{\mu\nu}=\sqrt{\tau_{b\star}}M_p F_{\mu\nu}
\end{align}
As mentioned, in order to integrate out the K\"ahler moduli $\tau_b$ and to understand the EFT of the axion and its quadratic interaction with the gauge field, a necessary condition is to have a hierarchy between the masses of the moduli and the axion such that $m_{\tau}\gg m_{\theta}$. In the case of LVS stabilization, the masses of the axions and the large cycle K\"ahler modulus are \cite{Conlon:2005ki}
\begin{align}
    m^2_{\theta_s}\sim \frac{M_p^2\ln ^2\mathcal{V}_\star}{\mathcal{V}_\star^2},\quad m^2_{\tau_b}=\frac{9|W_0|^2\hat{\xi}}{4 M_p^4 \mathcal{V}_\star^3}\sim \frac{M_p^2\ln ^{\frac{3}{2}}\mathcal{V}_\star}{\mathcal{V}_\star^3},\quad m^2_{\theta_b}\sim M_p^2 e^{-a_b\mathcal{V}_\star^{2/3}} ,
\end{align}
implying $m_{\theta_s}>m_{\tau_b}\gg m_{\theta_b}$. This tells us that the model admits the needed mass hierarchy between $\tau_b$ and $\theta_b$ in order to realize moduli backreaction. 
Using canonical normalization as outlined before,  the large axion decay constant $f_b$ is 
\begin{align}
    f_b=\sqrt{\frac32}\frac{M_p}{a_b\mathcal{V}_\star^{2/3}}.
\end{align} 

In this case, $m_{\tau_b}\gg m_{\theta_b}$, so we can integrate out the heavy moduli to develop an effective theory of $\theta_b$ and the gauge field only. 
From the LVS potential, Eq.~\eqref{eq: LVSPotential}, we can write the $\tau_b \theta_b^2$ interaction term as 
\begin{align}
    \mathcal{L}_{\tau_b \theta_b\theta_b}=-\frac{4}{3}\sqrt{\frac{2}{3}}\frac{a_b^3 A_b|W_0|(2+a_b \tau_{b\star})e^{-a_b \tau_{b\star}}}{M_p^5} \delta \hat{\tau}_b\hat{\theta}_b^2.
\end{align}
The tree level gauge kinetic term is
\begin{align}
     \mathcal{L}_{\tau_b \gamma\gamma}=-\frac{1}{2\sqrt{6}M_p}\delta\hat{\tau}_b \hat{F}_{\mu\nu}\hat{F}^{\mu\nu}
\end{align}
with effective coupling constant
\begin{align}
    g_{\theta_b\theta_b\gamma\gamma}=\frac{a_b A_b|W_0|(2+a_b\tau_{b\star})e^{-a_b \tau_{b\star}}}{3M_p^4\, m^2_{\tau_b}\, \tau_{b\star}^2} \frac{1}{f_b^2}
\end{align}
for large $a_b\tau_{b\star}$.
Consider for example ${\cal V}=10^3$, $a_b=0.1$, $A=M_p^3$, and $W_0=-M_p^3$. In this case we find that $g_{\theta \theta \gamma \gamma} \sim 10^{-2}/f^2 - 10^{-3}/f^2$ for $A_s=M_p^3$ and $a_s$ between $\mathcal{O}(10^{-1})-\mathcal{O}(1)$.  This is  significantly larger than the field theory benchmark from loops of charged pions. The coupling can be larger or smaller, depending primarily on the combination $a_b \tau_b$ due to the exponential suppression $e^{- a_b \tau_b}$.

\subsection{Non-Perturbative Corrections in Type IIB}

Finally we turn to non-perturbative corrections to the gauge kinetic function.

In Sec.~\ref{sec:origins}, we saw that there can be non-perturbative corrections to the gauge kinetic function and highlighted various sources of them. In order to write concrete couplings in KKLT and LVS like setups, we take a general structure of the gauge kinetic coupling as (note the same structure of the corrections across various setups \cite{Blumenhagen:2006xt, Akerblom:2007nh, Blumenhagen:2008ji, Blumenhagen:2009qh}) 
\begin{align}\label{eq: GauKinNP}
    f_{\mathrm{gauge}}=T+\frac{A^{\rm np}}{M_p^3}e^{-a_{\rm np} T}
\end{align}
where $T=\tau+i\theta$ and $A_{\rm np}$, $a_{\rm np}$ are parameters in the non-perturbative correction to the gauge kinetic function. As before, $\tau$ denotes the K\"ahler modulus of an internal 4-cycle which is wrapped by an ED3 instanton or a stack of D7-branes with gaugino condensate, and $\theta$ is the $C_4$ axion. The non-perturbative term could represent non-perturbative corrections from either type of source.

Consider first that the moduli are stabilized by a KKLT potential, as in Sec.~\ref{sec: KKLTbackreaction}, so the gauge kinetic function is given by equation (\ref{eq: GauKinNP}). The quadratic coupling to the axion comes from the expansion of the exponential in the non-perturbative contribution to the gauge kinetic function, giving
\begin{align}
   \mathcal{L}_{\rm quad,int}=  \frac{A^{\rm np}}{ M_p^3}\frac{a_{\rm np}^2}{12}\frac{\tau_{\star}}{M_p^2}e^{-a_{\rm np}\tau_\star}\,\hat{\theta}^2\hat{F}_{\mu\nu}\hat{F}^{\mu\nu}.
\end{align}
We write the coupling in terms of the axion decay constant as
\begin{align}
\label{eq:NPKKLT}
    g_{\theta\theta\gamma\gamma}=\frac{A^{\rm np}}{M_p^3} \frac{a_{\rm np}^2}{a^2}\frac{e^{-a_{\rm np} \tau_\star}}{8\tau_\star}\frac{1}{f^2}.
\end{align}
To estimate the coupling, consider for simplicity that $a_{\it np}=a$, and take the benchmark KKLT parameters, $a=0.1$, $A=M_p^3$, $\tau_*=113$. In this case we find $g_{\theta \theta \gamma \gamma}\sim 10^{-8}/f^2$. The coupling is exponentially sensitive to $a \tau_*$.

We can repeat this exercise in the context of LVS compactified on a Swiss-cheese Calabi-Yau threefold as in Sec. \ref{sec: LVSBackreaction}.  In this example, the stack of D7 branes wraps the big cycle.\footnote{If the D7 branes wrap the small cycle, we have a similar coupling between the small cycle axion and the D7 gauge theory on the small cycle. The treatment is similar.} The kinetic function of gauge fields living on the D7 branes receives a non-perturbative correction identical to the form Eq.~\eqref{eq: GauKinNP}, leading to a coupling of the canonically normalized fields given by

\begin{align}
    g_{\theta_b \theta_b \gamma\gamma}=\frac{A_b^{\rm np}}{M_p^3}\frac{a_{{\rm np},b}^2}{a_b^2}\frac{e^{-a_{{\rm np},b}\tau_{b\star}}}{4 \tau_{b\star}}\frac{1}{f_b^2}
\end{align}
expressed in terms of the big cycle axion decay constant $f_b$. This is identical to the KKLT result, Eq.~\eqref{eq:NPKKLT}, up to an ${\cal O}(1)$ constant.

This indicates that, in both KKLT and LVS, the quadratic axion coupling generated by non-perturbative effects has the general form,
\begin{equation}
    g \sim \frac{e^{ - a \tau_*}}{\tau_*} \frac{1}{f^2}
\end{equation}
and is therefore exponentially suppressed relative to $1/f^2$. Similar to the case of perturbative corrections, whether this is larger than the QCD axion coupling from charged pion loops depends on the precise details of the compactification via the parameters $a$ and $\tau_*$, as well as the parameters $a_{\rm np}$ and $A_{\rm np}$ which describe the non-perturbative effect.

\section{Discussion}
\label{sec:discussion}

In this work we have studied the string theory origins of a coupling between axions and the Maxwell field kinetic energy, namely interactions of the form ${\cal L}=\theta^2 F^2$ where $\theta$ is an axion-like particle and $F$ is the field strength tensor of a gauge field. Following a survey of possible mechanisms, we have identified specific paths within well studied corners of the string theory landscape, namely compactifications on Calabi-Yau orientifolds with moduli stablization achieved via KKLT or LVS. Through these analyses we have discovered many roads to the desired destination, which can broadly be classified as classical, perturbative, and non-perturbative contributions to the coupling. In KKLT and LVS models, we find $g_{\theta \theta \gamma \gamma} \ll 1/f^2$ for both perturbative and non-perturbative contributions (and classical couplings vanish in those cases). However, we note that the coupling can easily be larger than the loop-suppressed coupling of the field theoretic QCD axion, generated by loops of charged pions. These results, and the incredible sensitivity of atomic clocks and other experimental approaches to this portal, suggest that the quadratic axion coupling may provide an exciting new prospect for testing the axiverse.

There are many directions for future work. Foremost among these is detailed model building, namely the embedding of these mechanisms into explicit realizations of the Standard Model of particle physics in string theory. The gauge field of interest could instead be part of a dark sector, which motivates the further study of dark sector constructions in string theory. This also needs to be interfaced with axion model building, e.g., axion dark matter in string theory, including string theory realizations of the QCD axion. We leave these and other interesting directions to future work.

\acknowledgments

The authors thank the Banff International Research Station and the organizers of the workshop {\it Prospects for the String Axiverse} where this work was initiated. EM thanks Anirudh Prabhu and Joshua Foster for discussions at the BIRS workshop which motivated this work. RM is supported by the National Natural Science Foundation of China (NSFC) under Grant No. 12247103. NA and ARF are supported by NSERC via Subatomic Physics Discovery Grant 2026-00042. EM is supported by Research Manitoba via a New Investigator Operating Grant and by NSERC via a Subatomic Physics Discovery Grant 2022-00017.

\bibliographystyle{JHEP} 
\bibliography{refs}

\end{document}